# Review of Explainable Graph-Based Recommender Systems

THANET MARKCHOM, Department of Computer Science, University of Reading, United Kingdom
HUIZHI LIANG, School of Computing, Newcastle University, United Kingdom
JAMES FERRYMAN, Department of Computer Science, University of Reading, United Kingdom

Explainability of recommender systems has become essential to ensure users' trust and satisfaction. Various types of explainable recommender systems have been proposed, including explainable graph-based recommender systems. This review paper discusses state-of-the-art approaches of these systems and categorizes them based on three aspects: learning methods, explaining methods, and explanation types. It also explores the commonly used datasets, explainability evaluation methods, and future directions of this research area. Compared with the existing review papers, this paper focuses on explainability based on graphs and covers the topics required for developing novel explainable graph-based recommender systems.

## 1 INTRODUCTION

Due to the large exponential growth of information online, users are usually bombarded with an excessive number of content choices available. To cope with this information overload issue, recommender systems have become an essential tool to suggest pieces of information (items) that potentially match users' personal interests. During the early era of recommender systems, most were developed as black-box systems, meaning the internal mechanisms or decision-making processes were not transparent or easily interpretable [4]. Recent research has shed some light on potential issues that might be caused by using recommender systems without comprehension of how they generate outputs [7]. These issues include biases in making decisions and ethical violations, which can cause serious consequences [66]. To address such issues, recent regulations, such as the General Data Protection Regulation (GDPR) of the European Union and other countries [33], have been established. This emphasizes the importance of developing recommender systems that not only perform well in terms of accuracy but also provide explainability of recommendations [122].

Recognizing this necessity, numerous studies in recommender systems have focused on developing *explainable recommender systems* that are transparent and understandable. These systems aim to generate accurate recommendations while offering explanations for their recommendations or encouraging the predictability of explainable recommendations over non-explainable ones. These systems allow users to better understand the decision-making process, leading to increased trust and confidence in the systems. Also, by providing clear explanations for their recommendations, any biases that may be present in the data or algorithms can be identified and addressed accordingly. This ensures that the recommendations are fair, unbiased, and ethically sound.

To achieve both accuracy and explainability in recommendations, various resources have been utilized, including graphs. A graph is a collection of nodes (or vertices) and relations (or edges) connecting pairs of nodes. In recommendation scenarios, nodes can represent entities such as users, items, and attributes, while relations signify the connections between them, such as user-item interactions, social connections between users, and item attributes.

Authors' addresses: Thanet Markchom, Department of Computer Science, University of Reading, United Kingdom, thanet.markchom@reading.ac.uk; Huizhi Liang, School of Computing, Newcastle University, United Kingdom, huizhi.liang@ncl.ac.uk; James Ferryman, Department of Computer Science, University of Reading, United Kingdom, j.m.ferryman@reading.ac.uk.

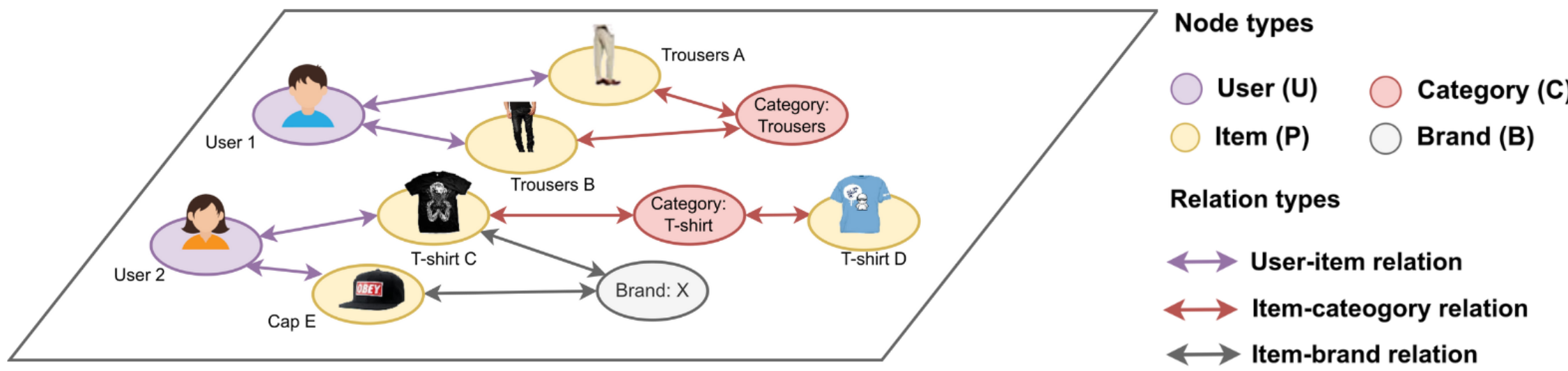

Fig. 1. Example of a graph in a clothing recommendation scenario

In recent years, graphs have been widely leveraged to develop *explainable graph-based recommender systems* [28]. These systems leverage graph structures to make personalized recommendations while also providing explainability of their recommendations via connectivity information within graphs. Unlike traditional recommender systems that predominantly rely on user-item interactions, explainable graph-based recommender systems leverage a variety of interconnected entities and relationships via multi-hop relations [98, 101, 108, 124]. A multi-hop relation represents a high-order connection between two nodes that are not adjacent. For example, in Figure 1, there is a multi-hop relation between "User 2" and "T-shirt D" through a path "User 2" - "T-shirt C" - "Category: T-shirt" - "T-shirt D". Based on this path, it is possible to recommend "T-shirt D" to "User 2" since they are linked through a high-order connection. Also, this multi-hop relation that connects them can serve as an explanation. This explanation can be interpreted as "T-shirt D" is recommended to "User 2" because it is in the same category as one of "User 2"'s previously bought items.

Explainable graph-based recommender systems represent a pivotal advancement in the recommendation research domain. They have garnered significant interest and have undergone continuous development [23, 35]. When developing an explainable graph-based recommender system, there are three principle aspects that should be considered: (1) a learning method, (2) an explaining method, and (3) an explanation type. A **learning method** pertains to the model architecture used to extract connectivity information within a graph for recommendation learning. Understanding these methods aids in the selection of a suitable architecture for developing a graph-based recommender system, taking into account the characteristics of available data and the methods used for extracting multi-hop relations. An **explaining method** refers to the mechanism of explainability component within a model. Understanding these mechanisms is crucial for incorporating explainability into graph-based recommender systems. An **explanation type** refers to the form of explainability. Choosing the appropriate type depends on stakeholder needs, with some scenarios requiring detailed node insights and others favoring higher-level path relationships.

Gaining insight into various learning methods, explaining methods, and explanation types can be highly beneficial for developing an explainable graph-based recommender system. Therefore, this review paper summarizes state-of-the-art explainable graph-based recommender systems and categorizes them based on these three aspects. Furthermore, this paper provides a summary of benchmark datasets and explainability evaluation techniques necessary for examining and validating the performances of explainable graph-based recommender systems. This will allow new researchers or anyone interested in explainable graph-based recommender systems to conveniently grasp the whole picture and identify the gaps in this research area.

**Contributions and Differences** The contributions of our review paper and the differences compared to other previously published review papers are summarized as follows: (1) This review focuses on explainable graph-based recommender systems unlike previous review papers that focus on graph-based recommend systems [28, 35, 107], explainable recommender systems encompassing various types, not exclusively graph-based ones [19, 122], explainable graph-based artificial intelligence [91] or explainable artificial intelligence [34, 66], which are broader topics compared to ours. (2) In this review, explainable graph-based recommender systems are categorized based on three aspects, i.e., learning method, explaining method, and explanation type. (3) In [23], the focus is on deep learning. In contrast, our paper covers both deep learning and machine learning systems. While [23] categorizes the systems based on how graphs, graph embeddings, or path embeddings are learned and merged with deep learning methods, our paper focuses on various aspects. Additionally, our paper discusses datasets and evaluation methods used in state-of-the-art systems. (4) In [16], the authors focused on visual explanations, exploring various types of explainable recommender systems broadly. In contrast, our paper reviews explainable graph-based recommender systems, covering aspects from their structure to the visualization of explanations. It provides a more in-depth focus on this specific type of recommender system.

The paper is organized as follows: Section 2 reviews definitions, notations, and the categorization of explainable graph-based recommender systems. Section 3 covers learning methods. Section 4 discusses explanation methods. Section 5 describes explanation types. Section 6 presents commonly used datasets, Section 7 reviews evaluation methods, and Section 8 concludes with future directions.

## 2 PRELIMINARIES

This section provides definitions, notations, and technical terms commonly used in explainable graph-based recommender systems. Then, it explains the categorizations of such systems.

### 2.1 Definitions, Notations, and Technical Terms

Definition 1. **(Graph)** In the context of recommendation, a graph (also known as a knowledge graph or an information network) is defined as a directed graph $\mathcal{G} = (\mathcal{N}, \mathcal{R})$, where $\mathcal{N}$ is a set of nodes (or entities), and $\mathcal{R}$ is a set of relations (or edges). Each node and relation is associated with its type mapping function: $\phi : \mathcal{N} \to \mathrm{N}$ and $\psi : \mathcal{R} \to \mathrm{R}$, respectively, where N represents the set of node types, and R represents the set of relation types. A triplet $(h, r, t)$ represents a relationship between the head node $h \in \mathcal{N}$ and the tail node $t \in \mathcal{N}$ connected by the relation $r \in \mathcal{R}$.

Definition 2. **(Path)** A path is a sequence of nodes where each adjacent pair is connected by a relation. Given a graph $\mathcal{G} = (\mathcal{N}, \mathcal{R})$, a path $(n_0, n_1, ..., n_{k-1}, n_k)$ exists if there exist triplets $(n_0, r_0, n_1)$, $(n_1, r_1, n_2)$, ..., $(n_{k-2}, r_{k-2}, n_{k-1})$, $(n_{k-1}, r_{k-1}, n_k)$ where $n_i \in \mathcal{N}$ for $i = 0, 1, ..., k$ and $r_i \in \mathcal{R}$ for $i = 0, 1, ..., k-1$ in $\mathcal{G}$. In other words, a path is a way to traverse a graph from one node to another by following the relations. The length of path is the number of relations it contains.

Definition 3. **(Single-hop relation)** Given a graph $\mathcal{G} = (\mathcal{N}, \mathcal{R})$, a single-hop relation is a direct relation between two nodes. A single-hop relation between nodes $n_0 \in \mathcal{N}$ and $n_1 \in \mathcal{N}$ exists if there exists a relation $r \in \mathcal{R}$ that connects them, i.e., there exists a triplet $(n_0, r, n_1)$ in the graph.

Definition 4. **(Multi-hop relation)** Given a graph $\mathcal{G} = (\mathcal{N}, \mathcal{R})$, a multi-hop relation refers to an indirect relation between two nodes through a path. Mathematically, a multi-hop relation between nodes $n_0 \in \mathcal{N}$ and $n_k \in \mathcal{N}$ exists, if there exists a path $(n_0, n_1, ..., n_{k-1}, n_k)$ such that $n_i \in \mathcal{N}$ for $i = 1, 2, ..., k$ in the graph.

Definition 5. **(Random walk)** A random walk on a graph $\mathcal{G} = (\mathcal{N}, \mathcal{R})$ can be denoted as a sequence of nodes $n_0, n_1, ..., n_k$ where $n_i$ is the node visited at step $i$ for $i = 0, 1, ..., k$. It is obtained through a stochastic process defined by the sequence of random variables $X_0, X_1, ..., X_k$, where each $X_i$ represents the node visited at step $i$. The walk starts at an initial node $n_0 \in \mathcal{N}$, and at each step $i$, the walker moves to a neighboring node $n_{i+1}$ of the current node $n_i$ with equal probability. The random walk continues indefinitely or until it forms a sequence of visited nodes of a certain length.

Definition 6. **(Meta-path)** [85] Given a graph $\mathcal{G} = (\mathcal{N}, \mathcal{R})$, a meta-path $\pi$, defined as $N_1 \xrightarrow{R_{N_1,N_2}} N_2 \cdots N_l \xrightarrow{R_{N_l,N_{l+1}}} N_{l+1}$ (abbreviated as $N_1N_2 \cdots N_{l+1}$), describes a composite relation $R_{N_1,N_2} \circ \cdots \circ R_{N_l,N_{l+1}}$ between $N_1$ and $N_{l+1}$ where $\circ$ denotes the composition operator on relations. A meta-path signifies a structural relationship among various types of nodes and relations in a graph. It serves as a tool to extract high-order connectivity information under a specific assumption within a graph. The length of meta-path $\pi$ is denoted by $|\pi|$.

Definition 7. **(Meta-path based random walk)** Given a graph $\mathcal{G} = (\mathcal{N}, \mathcal{R})$ and a meta-path $\pi = N_1N_2 \cdots N_{l+1}$, a meta-path based random walk is a sequence of nodes $n_0, n_1, ..., n_{l+1}$ where $n_i$ is the node visited at step $i$ for $i = 0, 1, ..., l+1$ such that $n_i$ has the type $N_i$ in $\pi$. It is obtained via a stochastic process on a graph, where the sequence of nodes visited is influenced by a predefined meta-path, capturing specific semantic relationships between nodes. The random walk continues until it reaches the last node type in the meta-path or continues indefinitely. In the case that it continues indefinitely, once it reaches the last node type in the meta-path, it goes back to the first node type and repeats following the meta-path again.

Definition 8. **(Node embedding method)** A node embedding method refers to the process of finding a mapping function $\Phi : \mathcal{N} \rightarrow \mathbb{R}^{|\mathcal{N}| \times d}$, where $d << |\mathcal{N}|$. This function maps a node into a latent space and generates a latent representation (embedding) of this node, capturing graph topological information. In a graph, node embeddings for all nodes can be obtained and utilized as input or features for learning recommendations.

Definition 9. **(Explainability and interpretability)** Explainable artificial intelligence (AI), including recommender systems, can offer explanations or help humans understand the logic behind its decisions. In this research area, two terms are usually mentioned, i.e., "explainability" and "interpretability". According to [7, 34], explainability refers to the ability to provide an explanation of why a decision is made. Meanwhile, interpretability refers to the characteristic of an AI system in which its internal mechanism is comprehensible to a human. For interpretable models, developers or users can leverage their interpretability to extract explanations. Therefore, whether an AI system is explainable or interpretable, it can provide explanations.

For instance, linear/logistic regression models are interpretable models because their coefficients directly indicate the magnitude and direction of influence that each corresponding input feature has on the output [12]. The relationship between input variables and the model's predictions makes these models interpretable and enables them to provide explanations for the predictions. On the other hand, neural networks are not interpretable due to their black-box internal mechanisms [34]. However, explanations can be derived from these networks by using techniques such as LIME [77], Anchors [78], SHAP [59], Global Sensitivity Analysis [21], ASTRID [40] or Concept Activation Vectors (CAVs) [46]. In other words, these methods allow neural networks to be explainable, providing them with the ability to offer explanations.

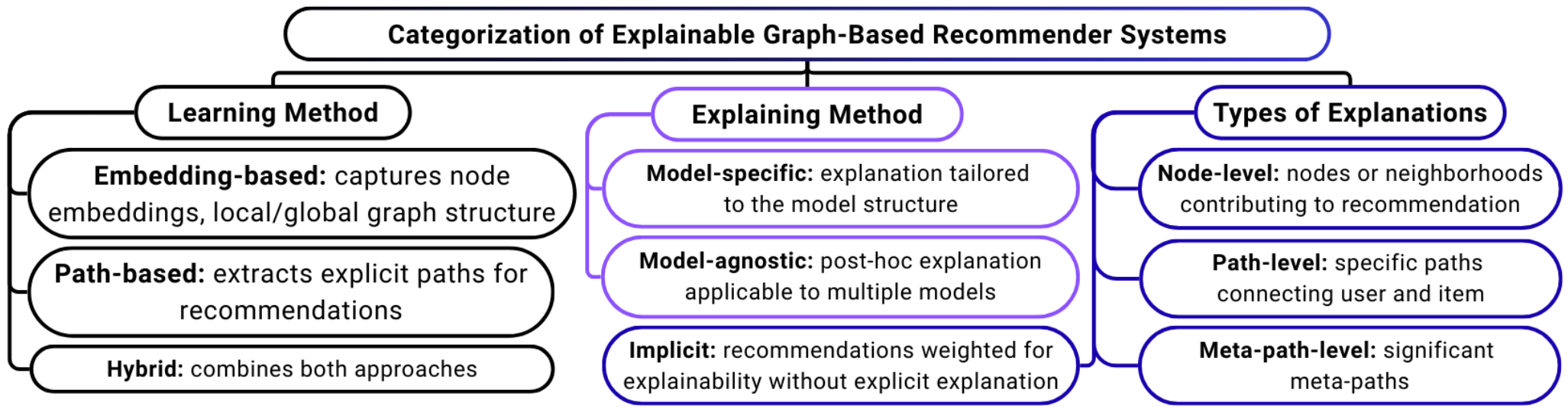

Fig. 2. Categorization of Explainable Graph-Based Recommender Systems

### 2.2 Categorization of Explainable Graph-Based Recommender Systems

Explainable graph-based recommender systems can be categorized by many aspects. In this paper, we consider three aspects, as shown in Figure 2. The details of these aspects are as follows:

(1) **Types of learning method (embedding-based, path-based, or hybrid)**: Graph-based recommender systems can be categorized by their learning methods into three categories, i.e., embedding-based, path-based, and hybrid approaches. In an embedding-based approach, node embeddings are utilized to capture either local neighborhood structures or global graph structures, without explicit consideration of the actual paths within the graphs. A path-based approach, on the other hand, explicitly extracts paths from a graph and incorporates these paths as information for learning recommendations. Lastly, a hybrid approach combines both path-based and embedding-based approaches to leverage the strengths of both methodologies.

(2) **Types of explaining method (model-specific or model-agnostic)**: Considering explaining methods, explainable graph-based recommender systems can be classified into two groups: model-specific and model-agnostic. Model-specific systems utilize explainability components that are tailor-made for their own structures and functionalities. In contrast, model-agnostic systems use post-hoc explainability components that can be applied not only to themselves but also to a wide range of other recommender systems.

(3) **Types of explanations (node-level, path-level, meta-path-level or implicit)**: Explainable graph-based recommender systems can further be categorized based on the formats of their explanations. These explanations can be in various forms, including node-level explanations (e.g., predictive or significant nodes, counterfactual nodes, or node neighborhoods), path-level explanations (e.g., selected paths connecting users and recommended items), and meta-path-level explanations (e.g., predictive or significant meta-paths). Additionally, there exists an implicit level where explicit explanations are not generated alongside recommendations. Instead, recommendations with higher explainability are emphasized over those with lower explainability in this category.

Table 1 summarizes state-of-the-art explainable graph-based recommender systems categorized by these three aspects. The details, including their learning methods, their explaining methods, and their explanation types, can be found in the following sections. The software packages for certain state-of-the-art explainable graph-based recommender systems are outlined in https://github.com/thanet-m/explainable-graph-based-recommender-systems.git.

| Model | Types of learning method | | Types of explainable model | | Types of explanation | | | |
|---|---|---|---|---|---|---|---|---|
| | Embedding-based | Path-based | Model-specific | Model-agnostic | Node | Path | Meta-Path | Implicit |
| SemAuto [9] | ✓ | | ✓ | | ✓ | | | |
| Ai et al.'s [5] | ✓ | | ✓ | | | ✓ | | |
| RippleNet [95] | ✓ | ✓ | ✓ | | | ✓ | | |
| SEP [113] | | ✓ | | ✓ | | ✓ | | |
| KGAT [98] | ✓ | | ✓ | | | ✓ | | |
| PGPR [108] | ✓ | | ✓ | | | ✓ | | |
| LDSDMF [6] | | ✓ | ✓ | | ✓ | | | |
| KPRN [102] | | ✓ | ✓ | | | ✓ | | |
| RuleRec [62] | | ✓ | | ✓ | | | ✓ | |
| EIUM [43] | ✓ | ✓ | ✓ | | | ✓ | | |
| Liu et al.'s [56] | ✓ | | ✓ | | ✓ | | | |
| HAGERec [114] | ✓ | | ✓ | | | ✓ | | |
| MSRE [103] | | ✓ | ✓ | | ✓ | ✓ | ✓ | |
| MP4Rec [69] | | ✓ | ✓ | | | | ✓ | ✓ |
| FairKG4Rec [27] | | ✓ | | ✓ | | ✓ | | |
| PRINCE [32] | | ✓ | | ✓ | ✓ | | | |
| ADAC [124] | ✓ | ✓ | ✓ | | | ✓ | | |
| CAFE [109] | ✓ | ✓ | ✓ | | | ✓ | | |
| GEAPR [50] | ✓ | ✓ | ✓ | | ✓ | | | |
| KGIN [99] | ✓ | | ✓ | | ✓ | | | |
| MKRLN [90] | ✓ | | ✓ | | | ✓ | | |
| TMER [17] | | ✓ | ✓ | | | ✓ | | |
| UCPR [88] | ✓ | | ✓ | | | ✓ | | |
| LOGER [127] | ✓ | | ✓ | | | ✓ | | |
| KGAT+ [81] | ✓ | | ✓ | | | ✓ | | |
| TPRec [126] | ✓ | | ✓ | | | ✓ | | |
| ReMR [101] | ✓ | | ✓ | | | ✓ | | |
| PLM-Rec [31] | | ✓ | ✓ | | | ✓ | | |
| METoNR [118] | | ✓ | ✓ | | | | ✓ | |
| KR-GCN [61] | ✓ | ✓ | ✓ | | | ✓ | | |
| CaDSI [100] | ✓ | ✓ | ✓ | | ✓ | | | |
| SEV-RS [64] | | ✓ | ✓ | | | | | ✓ |
| CGSR [117] | ✓ | | ✓ | | ✓ | | | |
| CGSR [106] | ✓ | | ✓ | | ✓ | ✓ | | |

Table 1. Summary of state-of-the-art explainable graph-based recommender systems categorized by three different aspects, i.e., types of learning method, types of explaining method, and types of explanations

## 3 CATEGORIZATION BY LEARNING METHODS: EMBEDDING-BASED, PATH-BASED, AND HYBRID APPROACHES

This section discusses the concepts, advantages, and disadvantages of different learning approaches and some state-of-the-art recommender systems within each approach.

### 3.1 Embedding-Based Approach

An embedding-based approach uses graph embedding methods [13] to generate node embeddings and incorporate them into recommendation frameworks. These embeddings capture the structural relationships among users, items, and other entities. This allows explanations to be generated by reasoning over the embeddings. Several methods can be applied to generate node embeddings, such as **autoencoders**, **translation-based models**, **reinforcement learning**, and **GNNs**. The following paragraphs discuss the use of each method.

**3.1.1 Autoencoder** An autoencoder is a neural network that learns efficient data representations by encoding inputs into a lower-dimensional space and reconstructing them. It is originally used for compression and dimensionality reduction. However, autoencoders can also be used for explainable graph-based recommender systems. Bellini et al. [9] used Semantics-Aware Autoencoders (SemAuto) [8] to incorporate structural information in a graph into an autoencoder. In SemAuto, each neuron corresponds to a graph node, and each connection represents an edge between nodes. The neuron weights are learned and used to construct user profiles, allowing the importance of each node for a specific user to be identified as an explanation.

**3.1.2 Translation-Based Models** Some embeddings-based models use translation-based methods such as TransE [11] and TransR [55] to generate node embeddings. Let $(h, r, t)$ be a triplet in a graph where $h$ denotes a head node, $r$ denotes a relation, and $t$ denotes a tail node. These methods are based on the assumption that the information from $h$ combined with the information from $r$ should equal the information from $t$, i.e., $h + r = t$. These embeddings have been proven to be highly flexible for many downstream tasks, including recommendation [95]. As a result, many systems have adopted this approach for developing explainable graph-based recommender systems. For instance, Ai et al. [5] proposed a method that models the distribution of node and relation embeddings by optimizing the generative probability of observed relation triplets similarly to TransE. Given a pair of a user and his/her recommended item, a breadth-first search is applied to find potential explanation paths. For each path, the probability based on the learned distribution of node and relation embeddings is computed. The path with the highest probability is applied to a predefined template in natural languages to generate a final explanation for the user. Zhu et al. [127] proposed LOGER, leveraging a neural logic model to guide the path reasoning process for generating explanations. A set of logical rules based on user-item interactions is first mined from a graph. Then, for each user, a set of personalized rule importance scores is generated by using Markov Logic Networks [74]. These scores, together with the node embeddings from TransE, are used to train an LSTM-based path reasoning network. Starting from a given user, this network predicts a sequence of nodes and relations until it reaches the recommended item, forming an explanation path.

**3.1.3 Reinforcement Learning Models** Finding an explanation path in a graph can be formulated as a deterministic Markov Decision Process (MDP). Thus, some embedding-based models use reinforcement learning (RL) to navigate paths connecting a user to a recommended item, with the traversed path serving as an explanation. For example, Xian et al. [108] proposed PGPR, using RL to navigate a graph to find recommended items. Generally, navigating through a graph requires a high amount of computational resources. To mitigate this, their method uses a user-conditional action pruning strategy to decrease the size of the action spaces. Moreover, since the policy-guided agent tends to repeatedly follow paths with the highest cumulative rewards, the resulting recommendations often lack diversity. To increase the diversity, a beam search algorithm guided by the policy network is applied to sample more diverse paths and rank them by their rewards. Tai et al. [88] proposed UCPR that considers both local and user-centric views. The local view is based on a sequence of actions taken by the agent, encoded using an LSTM. The user-centric view corresponds with the potential user's demand information in the path reasoning process. Assuming that a user's demand depends on a group of nodes and relations, not just a single node or relation, a user demand portfolio can be represented by a set of triplets $(h, r, t)$ where $h$ is a node connected with a given user at the $n$-th hop. These triplets are assigned different weights according to the user's demand. At each step, these weights are updated accordingly depending on the fulfilled demand in the previous step. The most relevant triplet is also identified during updating. The selected relevant triplets from every step are then finally formed into a reasoning path that reflects the highlighted triplets in the user demand portfolio.

Most RL-based methods typically ignored the interpretability of the generated explanation paths discovered by their path reasoning/finding processes. To mitigate this, Zhao et al. [124] proposed ADAC to ensure the interpretability of the generated explanation paths. The idea is to supervise the path finding process by using demonstrations, defined as

(1) shortest path, (2) meta-path based random walk, and (3) path of interest (random-walk paths in which most of the nodes match with users' interests). This method uses Generative Adversarial Imitation Learning [41] to create path and meta-path discriminators. These discriminators estimate how likely a discovered path follows the demonstrations. In this way, ADAC can learn to generate paths that are similar to the demonstrations to increase the interpretability of the explanations.

Some studies enrich graphs with additional information before applying RL to generate better recommendations and explanations. For example, Wang et al. [101] proposed ReMR that leverages both instance-level and ontology-level information in graphs. This method begins by constructing an ontology-view graph with the Microsoft Concept Graph and adding it to an original instance-view graph built from item metadata. This combined graph is then used in RL. Compared to existing systems that only consider an instance-view graph, ReMR can better capture users' interests and provide more diverse explanations with high-level concepts. However, adding an ontology-view graph expands the search space of RL, requiring more computational time. Zhao et al. [126] proposed TPRec, a method that leverages graphs augmented with temporal information. Their approach first extracts temporal statistical and temporal structural features from user-item interaction timestamps. These features are then projected into a temporal feature space and clustered using a Gaussian Mixture Model. Next, the user-item relations in the original graph are replaced with temporal relations corresponding to their clusters in the temporal feature space. TransE is then applied to generate node embeddings, followed by an RL-based model to learn path reasoning for recommendation. Tao et al. [90] proposed MKRLN to incorporate visual information in RL. This model utilizes a graph and images to generate state representations for the policy network. Based on these representations, the policy network learns to lead a user to his/her recommended item. Moreover, to reduce the size of action spaces, an attention mechanism is used to weight the importance of each neighbor.

Generally, using RL techniques to find reasoning paths for recommendations is a straightforward way to provide explanations. However, the major drawback of this approach is the scalability issue due to the size of the action spaces. These techniques require substantial computational resources, particularly for graphs with many nodes, relations, or high node out-degrees. How to effectively and efficiently find reasoning paths in a graph is still a challenging research problem for developing RL-based models.

**3.1.4 GNN-Based Models** Another line of embedding-based models uses Graph Neural Networks (GNNs) to access and leverage multi-hop information [107]. Through recursive propagation, node embeddings are updated based on their multi-hop neighbors, capturing high-order connectivity. Attention mechanisms are often used to identify nodes or relations that significantly influence user profiles and recommendations. For example, Wang et al. [98] proposed KGAT, an end-to-end model based on a Graph Convolution Network (GCN) and a Graph Attention Network [94]. First, KGAT employs TransR to generate initial node embeddings. Each embedding is then refined through recursive propagation of its neighbors' embeddings using a knowledge-aware attention mechanism. Given a user and his/her recommended item, the attention weights of cascaded propagation form explanation paths connecting them. Despite its effectiveness, this model suffers from high computational time, especially when the graph is large. Yang et al. [114] proposed HAGERec, which uses bi-directional information propagation to learn user and item embeddings. A hierarchical attention mechanism aggregates information from each node's local neighborhood, allowing significant neighbors to be identified and used as explanations.

The aforementioned models rely only on original information from nodes and relations in graphs to model users' profiles. To effectively profile users, some GNN-based models use higher concepts beyond nodes and relations in graphs to capture users' interests. Wang et al. [99] proposed KGIN to model users' underlying intents based on user-item interactions. Each intent is formed by attentively combining relation embeddings in a graph. The relation with the highest attention weight can be seen as an explanation of the user's intents. In [56], an improved aggregation method for GCNs was introduced. This method modifies the aggregation at the last layer of GCN to discriminate information from the lower layers based on concepts. Specifically, all embeddings from the lower layer are categorized into different concepts and then propagated separately. To categorize the embeddings into different concepts, the factor affiliation degrees are estimated from the learnable parameters of each factor. This allows the method to identify significant concepts for learning user and item embeddings and use them as explanations. Shimizu et al. [81] proposed KGAT+, an improved framework of KGAT that tackles the problem of scalability. The key idea is to compress one-to-many relations using a latent class model. First, a set of target relations for compression is defined. If relation $r$ in the triplet $(h, r, t)$ is a target relation, then a latent class between $h$ and $t$ is assumed. Then, the relation between "$h$ and $t$" is decomposed into "$h$ and the latent class" and "$t$ and the latent class" with certain probabilities. These probabilities are considered when computing the attention weights and loss for learning recommendations. According to their experiments, KGAT+ required less computational cost and produced recommendations with higher interpretability compared to KGAT.

Some recent GNN-based models incorporate causality into graphs to enhance both recommendation quality and explainability. For instance, Yu et al. [117] proposed CGSR, which blocks shortcut paths on the session graph while retaining crucial causal relations for learning users' preferences. Shortcut paths are defined as direct links between two items that disregard sequential dependencies and intermediate interactions. CGSR consists of two main components: distillation and aggregation. In the distillation component, back-door adjustment of causality [20, 72, 116] is used to block shortcut paths in the session graph, resulting in a distilled session graph that preserves causal relations among items. The aggregation component then applies GNNs to aggregate high-order information from different relation types on the distilled session graph. This approach enables CGSR to learn user and item representations that reflect the causal relationships within the graph. Wu et al. [106] proposed a method, also called CGSR. Their method begins by constructing three types of graphs: cause, effect, and correlation graphs. The cause and effect graphs are built based on causal relations, specifically the cause and effect relationships between items. The correlation graph captures first-order relationships in the session graph as well as various second-order relationships derived from it. After that, a GNN with an attention mechanism generates item embeddings from each graph and aggregates them to produce final session representations for learning recommendation.

Although GNN-based methods have been widely used to develop explainable graph-based recommender systems, several challenges remain. One major issue is the long-tail distribution of node degrees, which results in many cold users and items. Capturing the structural information of these cold users and items often requires deeper GNNs. However, increasing the number of layers can introduce noise during propagation [48], potentially compromising both recommendation accuracy and explainability. Therefore, determining the optimal number of GNN layers to balance performance and interpretability remains an open research question [57, 107].

### 3.2 Path-Based Approach

A path-based approach typically extracts user-item paths and feeds them to a recommendation learning framework to learn the connectivity possibilities between users and items. Depending on path extraction methods, a path-based approach can be categorized into two groups: (1) **random-walk based models** that extract or sample paths randomly before using these paths for learning recommendations and (2) **meta-path based models** that use paths extracted via meta-paths [85] to feed the recommendation frameworks.

**3.2.1 Random-Walk Based Models** The models in this group use random walks [58] to sample paths, starting from a node and randomly selecting adjacent nodes until a path reaches a set length or meets certain conditions. Then, the sampled paths are fed into a recommendation framework. For example, Wang et al. [102] proposed KPRN, which uses an LSTM to capture sequential dependencies of nodes and relations in randomly sampled paths. Each path is encoded into a representation, and all path representations are combined via attention-weighted pooling. The aggregated representation is fed to fully connected layers to compute a recommendation score, with attention weights highlighting the most influential path as an explanation. Geng et al. [31] utilized sampled paths to build a path language model for recommendations called PLM-Rec. This model generates candidate path instances connecting a user node to a recommended item node by using a Transformer-based decoder, similar to an auto-regressive language model [22]. Then, the recommendation scores are computed by the joint probabilities of these candidate path instances.

With a random walk, there is a high chance that nodes with higher degrees will appear more in the extracted paths than those with lower degrees. This may result in obtaining paths with low diversity, causing some biases in recommendations and explanations. To address this issue, Fu et al. [27] proposed FairKG4Rec, a method designed to produce fair explanation paths. Given pre-selected user–item paths extracted by any model, FairKG4Rec incorporates fairness considerations when re-ranking and selecting explanation paths. This method quantifies fairness using path and diversity scores. The path score indicates the quality of paths based on user–item historical data. It deals with the bias of user–item path patterns in historical records. The diversity score is based on Simpson's Index of Diversity of candidate explanation paths. This score can be considered as the regularization for improving explanation path diversity.

Apart from diversity issues, using paths extracted via a random walk method may breach a privacy issue since these paths may contain other users' historical data. Ghazimatin et al. [32] proposed PRINCE to address privacy concerns in a path-based approach. This model provides an explanation in the form of a set of minimal actions performed by the user that can change the recommendation to a different item (counterfactual). First, the Personalized PageRank (PPR) [37, 68] model, which uses a biased random walk, generates recommendations based on a graph. Given a user, a pair of recommended items is obtained from the PPR model. Then, each outgoing edge of this user is iteratively removed to examine whether its removal alters the ranking order of the recommended items. Finally, the edges whose removal changes the ranking are considered counterfactual explanations for the user.

Some methods systematically explore paths in a graph in a deterministic manner, rather than randomly sampling them. Instead of performing stochastic random walks, they follow all possible paths up to a certain length, resulting in an exhaustive traversal while still allowing all nodes and relations to be considered. This process can be interpreted as a random walk in which every outgoing edge has probability 1. For instance, Alshammari et al. [6] developed LDSDMF, which uses all 1-hop and 2-hop paths to measure item-item similarity and add explainability to a black-box MF model. This method first constructs a semantic graph from items and their attributes extracted from DBpedia. It then computes similarity between items using the linked data semantic distance (LDSD), considering either direct 1-hop links or 2-hop paths. An explainability regularization term based on this similarity is incorporated into the MF loss function, constraining the latent factors of positive and negative items. Yang et al. [113] proposed SEP, a post-hoc method to select explanation paths for any recommendation from a graph. Given a user–item pair, SEP uses a revised depth-first search algorithm [79] to identify candidate paths connecting them in the graph. These paths are ranked according to three heuristic metrics: (1) path credibility, computed as the product of all weights along the path, (2) path readability, inversely proportional to the path length and the number of intermediate node types, and (3) path diversity, reflecting the variety of relation types. The metrics are combined via a weighted sum, and the scoring function is learned using an unsupervised method.

**3.2.2 Meta-Path Based Models** Random-walk-based models sample paths probabilistically, rendering the connectivity information implicit and less predictable. To obtain such information in a more explicit and semantically meaningful manner, meta-path based random walks were proposed [85]. This technique has been widely adopted to extract paths under specific assumptions. Instead of randomly choosing the next node, meta-path based random walk selects the next node according to the node type specified by a given meta-path. Some examples of meta-path based models are discussed here. Ma et al. [62] proposed RuleRec, which uses meta-paths derived from item associations. It identifies item-to-item meta-paths through hard-selection (using a predefined threshold) or soft-selection (learning to select meta-paths via optimization) methods. Then, given a pair of items, path instances connecting them are extracted based on the selected meta-paths. Probabilities of path instances following these meta-paths are computed to form feature vectors of the items, where each entry represents the probability corresponding to each meta-path. These vectors are then used for learning recommendations with BPR-MF [76] and NCF [39] models. Chen et al. [17] developed TMER, leveraging paths connecting consecutive items in each user's history and attention mechanisms. Given a user-item pair, this method uses DeepWalk [73] to generate initial user and item embeddings. Then, for each pair of consecutive items in this user's history, multiple item-to-item meta-paths are used to extract path instances connecting them. Word2Vec [65] is then used to compute path embedding. This path embedding is attentively combined with an initial item embedding to generate a new item embedding using a self-attention mechanism. The attention weights are used to find reasoning paths as explanations for the user. Wang et al. [103] proposed MSRE, which uses two structural relations, i.e., affiliation relation and interaction relation, based on meta-paths. For the affiliation relation, the meta-paths following user-item-features and item-user-attributes patterns are used. For the interaction relation, meta-paths starting with a user node type and ending with an item node type are used. Moreover, to handle the large number of possible meta-paths, MSRE samples only those with high association degrees. Based on these sampled meta-paths, meta-path embeddings are computed and attentively aggregated via an attention mechanism. The most important meta-path is identified by the attention weights and used as an explanation.

Ozsoy et al. [69] propose MP4Rec, which uses PathSim [86] and metapath2vec [24] to compute user–user and item–item similarity matrices along different meta-paths. From these matrices, a Multi-Layer Perceptron (MLP) learns latent factors for each meta-path. The latent factors from all meta-paths are then aggregated via an attention mechanism to produce the final user and item embeddings, which are used in a BPR framework. The most important meta-path, indicated by attention weights, serves as an explanation. To further enhance explainable recommendations, MP4Rec adds a constraint to the BPR loss. This constraint is based on the explainability of user–item pairs, calculated using association rule mining [70], and encourages explainable items for each user to be close to that user in the latent space. Following a similar idea to [69], Markchom et al. [64] proposed SEV-RS, which addresses the scalability issues of using meta-paths. SEV-RS uses a scalable method to extract meta-path based multi-hop information (used as recommendation-learning features) and compute meta-path based explainability scores for any user–item pair. These extracted features and explainability scores are then incorporated into a BPR framework to learn and constrain explainable recommendations, similar to [69]. Zhang et al. [118] proposed METoNR, which uses meta explanation triplets, each represented by a symmetric 2-hop meta-path, to provide explanations. In this method, initial news representations are learned from titles using a Transformer, and user preference representations are extracted from interaction sequences using a GRU network. Side information for news and users is derived from meta explanation triplets to capture graph-based semantic connections. Multiple side information representations based on different triplets are combined via a self-attention mechanism to identify the most important triplet for explanation. Finally, initial and side information representations are concatenated to form the final embeddings.

Evidently, the effectiveness of meta-path based models highly depends on selecting meta-paths that are suitable for the target domains. This selection usually requires domain knowledge and can be labor-intensive, particularly for graphs with diverse node and relation types [98]. Some studies also argued that meta-path based models are inefficient since they are not capable of utilizing unseen connectivity patterns that are not specified in given meta-paths [61, 102]. These observations indicate that there is still a need for further investigation into meta-path based explainable recommender systems.

### 3.3 Hybrid Approach

An embedding-based approach uses latent node representations that capture both local and global views of the graph structure. It offers greater flexibility and efficiency than a path-based approach. A path-based approach, in contrast, uses explicit paths containing high-order information. This provides a more intuitive and interpretable way of leveraging multi-hop connectivity within graphs compared to a node-embedding approach. A hybrid approach combines both approaches. It leverages embeddings for local and global graph structure, and incorporates extracted paths for richer context and higher interpretability. For instance, Wang et al. [95] introduced RippleNet, which uses ripple sets, triplets within $n$ hops from a user's interacted item. The method first generates ripple sets for each user–item pair at different hop levels. These sets correspond to multi-hop propagation paths centered on a given user. Starting from the 1-hop set, information from each triplet is propagated based on a relevance probability defined as the similarity between the item embedding and the head node embedding. The user embedding is then computed as a weighted sum of the tail node embeddings using these relevance probabilities. This process is repeated across all $n$-hop ripple sets to produce a final user embedding enriched with multi-hop graph information. The resulting user and item embeddings are used to predict recommendation scores. Explanations are obtained by tracing a path from the user to the recommended item, selecting at each hop the node with the highest relevance probability.

Huang et al. [43] proposed EIUM, a model that jointly leverages multi-hop graph relations together with textual information from user reviews and visual information from item images. The model first applies TransE, constrained by multimodal features, to generate node embeddings. Next, a set of user–item paths is extracted. For each path, the node embeddings from the constrained TransE are used to compute a path representation. All path representations are then combined through an attention mechanism to form the final user–item interaction representation, with explanations provided by selecting the path instance that receives the highest attention weight. Li et al. [50] proposed GEAPR, a multimodal POI recommender system that integrates graph structure, neighbor influence, user attributes, and geolocation. The system captures structural context using random walk with restart (RWR) [92] combined with a neural network, neighbor influence through an attention-based GNN, and user attributes through an attention-based latent factorization machine. All features are fused using the attention mechanism from [49], which enables using attention weights as explanations.

Xian et al. [109] proposed CAFE, a coarse-to-fine neural symbolic reasoning method. This method forms user profiles as coarse sketches of behavior to guide path finding for explanations. User-centric patterns (defined as relational paths reflecting personal interests) are first identified using a random walk, and prominence scores are computed to combine them into a user profile. Using this profile, the Profile-guided Path-Reasoning (PPR) algorithm guides neural symbolic reasoning modules in finding explanation paths using a layout tree of relation types. Compared to RL-based models, CAFE is more efficient since it can simultaneously find multiple explanation paths rather than finding each path separately. Ma et al. [61] proposed KR-GCN, a path reasoning model with a GCN. The model uses a GCN to learn node representations, where a weighted-sum aggregator combines each node with its neighbors, followed by a non-linear activation. Path instances are extracted through a heuristic path search guided by TransH scoring and nucleus sampling to filter out low-quality paths. The selected paths are then encoded with an LSTM and a self-attention mechanism based on the node representations from a GCN. The attention scores highlight the most important path, providing interpretability. Finally, multiple MLP layers are applied to the encoded vector to generate recommendations.

Wang et al. [100] introduced CaDSI, using a causal graph to provide interpretable recommendations. Their method begins with meta-path based random walks and metapath2vec to generate semantic-aware embeddings for users, items, and aspects. From these embeddings, a GNN-based disentangling module is then applied to learn user representations that capture multiple user intents. After obtaining these representations, the backdoor adjustment technique [71] is applied to produce unbiased user representations. These unbiased intent-aware user representations are then used to compute recommendation scores with a second-order Factorization Machine [75].

### 3.4 Summary

**Embedding-based models** represent entities and relations as dense vectors that capture complex semantic patterns and generalizing across graph structures. They offer flexibility but can be computationally expensive, hard to interpret, and sensitive to sparse or noisy data. **Path-based models** use explicit paths to infer recommendations. They provide more transparent, interpretable results and can be more scalable than embedding-based models. However, they may miss global patterns, rely on predefined paths, and produce inaccurate results if the paths used are of low quality. Using excessively long paths can increase the time complexity of path extraction [51] and may introduce noise [86], whereas limiting path length can be sufficient for effective recommendation learning [86, 87]. **Hybrid models** combine embedding-based and path-based approaches to leverage both global structural information richness and interpretability. They offer flexibility and explainability, but are more complex to design. The choice of approach depends on graph characteristics, recommendation goals, and the trade-offs between accuracy, interpretability, and efficiency.

## 4 CATEGORIZATION BY EXPLAINING METHODS: MODEL-SPECIFIC AND MODEL-AGNOSTIC APPROACHES

Explainable graph-based recommender systems can also be categorized into two approaches based on how their explainability mechanisms work: **model-specific**, where the mechanisms are developed specifically for the models they are in, and **model-agnostic**, where the mechanisms are also applicable to other recommender systems. This section discusses the characteristics and differences of these approaches, the methods used in each, and their respective advantages and disadvantages.

### 4.1 Model-Specific Approach

The model-specific approach uses an internal component/mechanism within a recommendation model to provide insights into the decision-making process specific to that model. Various techniques have been employed to develop model-specific explaining methods for explainable graph-based recommender systems. Notable among these techniques are **attention mechanisms**, **reinforcement learning**, and **auto-regressive path generation**. In the following paragraphs, the details of these techniques are discussed, as well as their state-of-the-art applications.

**4.1.1 Attention Mechanism** As in the previous section, attention mechanisms have been heavily used to signify the importance of nodes [50, 114] or relations [8, 81, 98, 99] and use them to produce explanations. The models using these mechanisms are considered a model-specific approach, as the attention weights are learned jointly with the internal components that produce the recommendations.

Attention mechanisms have been applied in many model architectures, including GCNs. For example, both KGAT [98] and KGAT+ [81] use the knowledge-aware attention mechanism on GCN models to calculate the attention weights of a given user's neighbors. The weights are learned by optimizing the TransR loss for capturing the graph structure and the collaborative filtering loss for learning recommendations. HAGERec [114] uses a bi-directional information propagation method with a hierarchical attention mechanism to learn the importance score of each neighbor during propagation. GEAPR [50] uses three different attention mechanism modules to attentively aggregate multimodal information, i.e., the structural context of a user, numerical user's attributes, and categorical user's attributes. Structural features are aggregated by an attention-based GNN. The most important neighbor can be identified from the attention weights. Then, latent factors of both numerical and categorical users' attributes are extracted by an attention-based latent factorization machine. Finally, these multimodal extracted features are combined by using another attention mechanism. KGIN [99] also uses an attention mechanism to combine relation embeddings to form each user intent representation. This allows the relation with the highest attention weights to be used as an explanation.

Besides identifying the most important node/relation, some models further take advantage of attention weights to derive explanation paths or meta-paths. MSRE [103] employs attention-guided walks to discover multi-style explanation meta-paths based on affiliation and interaction relations. TMER [17] applies an attention mechanism to identify paths linking consecutive items in a user's history, using them as explanations for recommendations.

Instead of applying attention mechanisms to nodes or relations, some models focus on a set of extracted paths, attentively aggregating their information to identify the most important path and use it as an explanation. For example, KPRN [102] samples paths connecting a user-item pair, encodes them with an LSTM network, and applies an attention mechanism to aggregate the path representations. EIUM [43] uses self-attention to learn path representations from user and item representations within the paths, then applies another attention mechanism with weighted pooling to combine them and identify the most important path as the explanation. KR-GCN [61] also uses an attention mechanism to distinguish the contributions of selected paths from its heuristic path extraction method. Attention mechanisms can also be applied on a meta-path level. MP4Rec [69] initially generates user and item latent factors based on various meta-paths. Then, it uses an attention mechanism to attentively aggregate information based on different meta-paths.

Some models compute the importance of nodes or relations during recommendation learning, in a manner similar to the attention mechanisms discussed previously. For example, SemAuto [8] transforms the graph into a neural network where each relation becomes an edge, and the learned neuron weights from the optimization process represent the significance of the corresponding relations. These weights identify the key attributes of a user as explanations. In [56], the modified aggregation method in a GCN propagates low-level embeddings based on computed factor affiliation degrees. These degrees are calculated from the learnable parameters of their model. They allow significant concepts for learning user and item embeddings to be identified and used as explanations. In CaDSI [100], importance scores are computed using an interactive update rule for each user-item interaction under a specific intent, allowing the model to explain recommendations based on the intent that contributed most to the prediction. CGSR [106] generates explanation scores at both session and item levels, revealing the session's overall impact on the recommended item and the importance of each item in the session as a cause or correlation.

#### 4.1.2 Reinforcement Learning

Another line of work in a model-specific approach uses RL for path reasoning, learning to navigate from a user to a recommended item. Various strategies for this navigation have been proposed. For example, PGPR [108] uses a multi-hop scoring function based on high-order connectivity between users and items to reward the RL agent. To improve efficiency, a user-conditional action pruning strategy is applied to reduce the size of the action space. UCPR [88] uses an LSTM network to encode a sequence of actions taken by the agent and a user-centric view reasoning network to construct and update each user's profile accordingly. ADAC [124] leverages Generative Adversarial Imitation Learning [41] to guide the path reasoning process to generate explanation paths following the path demonstrations. ReMR [101] uses a Cascading Actor-Critic method to initially learn the policies based on high-level concepts of users' personal interests. Then, these policies are used to learn the more fine-grained concepts to specify users' interests at an instance level. For each user and his/her recommended item, an explanation path is extracted using a multi-level reasoning path extraction method consisting of different levels of concepts and instances. TPRec [126] uses RL for path reasoning with a personalized time-aware reward function, replacing the multi-hop scoring function [108]. Path reasoning using RL can also leverage attention mechanisms. For example, MKRLN [90] uses an attention mechanism to compute the importance of neighbors at each navigation step. This not only improves navigation effectiveness but also reduces the number of action spaces by filtering out neighbors with low attention weights.

#### 4.1.3 Auto-Regressive Path Generation

Some model-specific explainable graph-based recommender systems generate explanation paths instead of extracting them. For instance, LOGER [127] uses an LSTM to predict a sequence of nodes and relations from a user representation, forming a path to the recommended item. PLM-Rec [31] employs a Transformer-based decoder to generate candidate paths and selects the one with the highest joint probability as an explanation. CAFE [109] uses neural symbolic modules and a layout tree to generate multiple explanation paths simultaneously, offering greater efficiency than models that generate paths one by one.

#### 4.1.4 Others

Other model-specific explanation techniques include learning from graph structures or causal relationships. In [5], node and relation representations were learned from graph triplets, and candidate paths extracted via breadth-first search were ranked using these representations. CGSR [117] constructs a distilled session graph by removing shortcut paths and retaining only causal item relationships. Item representations derived from this causal context produce similarity scores that reflect underlying causal connections. For each user, these scores provide explanations based on causality-driven similarities with his/her previously interacted items.

### 4.2 Model-Agnostic Approach

Unlike model-specific methods, model-agnostic methods can be applied to any recommendation model. They typically extract candidate attributes, such as nodes, relations, or paths, and rank or validate them using specific criteria. For example, [6] generates explanations based on a user's likability of an item attribute, determined by how often the user interacts with items possessing that attribute. SEP [113] selects an explanation by ranking candidate paths extracted from a graph. The ranking criteria are based on three heuristic metrics, i.e., path credibility, path readability, and path diversity. The final ranking score can be computed by the weighted sum of these metrics. Another post-hoc path selection method was proposed in [27]. Unlike other methods, this method takes fairness into account when ranking candidate paths. PRINCE [32] uses counterfactual explanations instead of explanation paths. Given a user and a pair of his/her recommended items, this method iteratively removes one of the relations from this user in a graph to check whether it alters the order of the given recommended items. If it changes the order, then this relation may have a high impact and can be used as a counterfactual explanation for this user.

Some model-agnostic methods extract multi-hop features to improve recommendation explainability, with these features being applicable across different recommendation frameworks. For instance, RuleRec [62] uses pre-defined meta-paths to create item feature vectors, where each entry reflects the connectivity probability between users and the item along a meta-path. These features can be incorporated into various recommendation models, such as MF-based models, using a multi-task learning objective that optimizes both feature extraction and recommendation simultaneously, making this method model-agnostic.

### 4.3 Summary

**Model-specific methods** are integrated into a particular recommender system, offering tailored and precise explanations aligned with its recommendations. However, they may lack generalizability, increase system complexity, and require trade-offs between accuracy and explainability. **Model-agnostic methods** can be applied post-hoc to various models, providing flexible and transferable explanations. They may be less precise and more generic compared to model-specific methods. The choice depends on the desired balance between specificity, generalizability, and system requirements.

## 5 CATEGORIZATION BY TYPES OF EXPLANATIONS: NODE, PATH, META-PATH, AND IMPLICIT LEVELS

This section centers on categorizing explainable graph-based recommender systems according to different explanation types. These types include **node-level, path-level**, **meta-path level**, and **implicit** explanations. The details of each type are discussed in this section with examples from recently proposed explainable graph-based recommender systems. Additionally, a discussion on the advantages and disadvantages of each explanation type is provided.

### 5.1 Node Level

Node-level explanations are the simplest explanation type, typically consisting of nodes or relations that significantly influence a user's recommendations. For example, SemAuto [9] treats the neuron weights in Autoencoder Neural Network as the importance weights of nodes. Then, it identifies the most important node connected to a given user and uses it as an explanation. GEAPR [50] employs a GNN with attention to aggregate information from a user's neighborhood, allowing the most important neighbor node to be identified based on attention weights. LDSDMF [6] identifies item attributes that match a user's preferences, providing explanations as the attribute node with the highest likability degree. KGIN [99] models user-item interaction intents using graph relations, combining them with attention to form intent representations, with the highest-weighted relation serving as the explanation. The model in [56] assigns user and item nodes to concepts representing semantic factors (can be considered as nodes). A modified GCN propagates neighbor information according to these concepts, and explanations are provided as the most significant concepts for each user or item. CaDSI [100] provides explanations in the form of user-item interaction relations with important scores based on certain specific intents. CGSR [106] provides importance scores of items to explain session-based recommendations, indicating item nodes that are significant to the recommended item. CGSR [117] computes similarity scores between connected item nodes, allowing a recommendation to be explained by the previously purchased item most similar to it. Node-level explanations can also include counterfactual nodes or relations. For example, PRINCE [32] provides counterfactual explanations by identifying relations that, if removed, would change the recommendation outcomes, representing conditions that oppose the actual results.

### 5.2 Path Level

Paths containing high-order connectivity information are typically extracted or considered to model users' profiles and learn recommendations in graph-based recommender systems. Using them as explanations is a straightforward way to utilize such information uniquely provided by graphs. According to this, many explainable graph-based recommender systems have been focusing on providing path-level explanations. SEP [113] and FairKG4Rec [27] extract candidate paths connecting a user and his/her recommended item and rank them based on certain metrics and conditions. CGSR [106] provides session-level scores that indicate the impact of a specific session, represented as a path, on the recommendation. Path-level explanations can also be modified or rewritten to obtain more user-friendly explanations than sequences of nodes and relations. For instance, the model in [5] ranks candidate paths using probabilities derived from node embeddings learned by TransE, then selects the highest-probability path and applies it to a predefined natural language template.

Besides defining criteria or metrics to measure the significance of candidate paths, some models, such as KPRN [102], EIUM [43], and TMER [17], use attention mechanisms to attentively combine multiple paths and identify the most significant one from the attention weights. KGAT [98], KGAT+ [81], and HAGERec [114] use attention mechanisms to determine node importance, enabling the construction of explanation paths for a user-item pair by sequentially following the attention weights. RippleNet [95] similarly leverages the relevance probabilities to form an explanation path.

Explanation paths can also be generated based on the distribution of nodes in a graph, as in an auto-regressive path language model in which sentences are formed based on the distribution of words in a corpus. For example, LOGER [127] uses an LSTM-based model to generate explanation paths. PLM-Rec [31] generates explanation paths by using a Transformer-based decoder. CAFE [109] uses multiple neural symbolic modules to first generate a tree where the root node is a user node and the leaf nodes are the recommended item nodes. Based on this tree, multiple explanation paths can be extracted by tree traversal and ranked to find the most suitable one.

PGPR [108], ADAC [124], UCPR [88], MKRLN [90], TPRec [126] and ReMR [101] formulate a recommendation task as deterministic MDP of path reasoning. The task is to navigate from a user node to his/her recommended item node, forming an explanation for this recommendation. One advantage of these models is that paths discovered by a path reasoning process definitely exist in a graph since they are generated by navigating through an actual graph. This is different from paths generated from node distributions, which may not exist in a graph.

### 5.3 Meta-Path Level

Path-level explanations provide explainability at an instance-level view. On the other hand, meta-path level explanations provide explanations from a higher meta-level view. These explanations are in the form of node types or relation types in graphs. They can be used to explain recommendations based on the semantic meanings of these meta-paths. MSRE [103] leverages several meta-paths to extract multi-hop relations. A meta-path embedding method aggregates information from these meta-paths by using an attention mechanism. The most important meta-path can be identified based on the attention weights and used for explaining the recommendations. Similarly, MP4Rec [69] models user and item latent factors using multiple meta-paths. First, the initial user and item latent factors are learned from the corresponding user–user and item–item similarity matrices derived from each meta-path. Then, the latent factors obtained from different meta-paths are combined through an attention mechanism, where the most significant meta-path can be identified from the attention weights. RuleRec [62] uses hard- and soft-selection methods to choose item-to-item meta-paths for generating item features. These features are used in a recommendation model, and the selected meta-paths provide explainability for the resulting recommendations.

Note that meta-path based models in Section 3.2.2 are recommendation models that utilize meta-path based random walks for extracting and leveraging connectivity information in a graph. Meanwhile, meta-path level categorization in this section is intended for any graph-based recommender systems that provide explanations in the form of meta-paths. These systems can employ various techniques and are not constrained to rely on meta-path based random walks, unlike meta-path based models.

### 5.4 Implicit

Some models provide explainability implicitly by selectively recommending items that are explainable rather than non-explainable ones. The methods in [69] and [64] uses a modified BPR framework to increase the chance of generating more explainable items rather than those non-explainable ones. Specifically, these methods add an additional term in the classic BPR loss function. This term constrains the distance of the user and item latent factors in the latent space to be close to each other (i.e., their difference is close to zero) if the item is explainable to the user.

### 5.5 Summary

**Node-level explanations** highlight the importance of individual nodes, making recommendations easy to interpret and computationally inexpensive. However, they lack context about relationships between nodes and may not capture the system's overall behavior. **Path-level explanations** show sequential relationships between nodes, offering context and insight into user behavior. They are more complex and computationally demanding, which may reduce efficiency and scalability. **Meta-path level explanations** provide higher-level abstractions suitable for diverse recommendation scenarios. They may lose fine-grained details and introduce additional complexity and computational overhead. **Implicit explanations** avoid explicit details, maintaining a clean interface and uninterrupted user experience. However, they reduce transparency, making it harder for users to understand recommendations and for developers to debug or improve the system. Choosing an explanation level depends on the required depth of insight, computational resources, and user needs. Node-level explanations are simpler and faster, while path- or meta-path level explanations provide richer context, and implicit explanations prioritize interface simplicity. Balancing these factors ensures both effective recommendations and user satisfaction.

## 6 BENCHMARK DATASETS

Benchmark datasets are crucial for advancing research in explainable graph-based recommender systems. This section reviews the datasets used in existing studies, detailing their characteristics, variations, applications, and popularity. Several datasets commonly used in the previous work are summarized in Table 2. The details of these datasets are as follows:

**MovieLens**[1] is a widely used benchmark dataset in movie recommendations. It contains user records of both explicit feedback (rating) and implicit feedback (watching and tagging). Regarding items, this dataset contains metadata of movies such as genres, directors, actors, tags, etc. There are multiple versions of this dataset with different scales, including MovieLens-HetRec-2011, Movielens-100K [6], Movielens-1M [88], and Movielens-20M [43]. Moreover, to incorporate richer information about movies, the MovieLens dataset can be combined with external resources. For example, in [102], MovieLens-1M was linked with the IMDb dataset using titles and release dates. In [43], MovieLens-20M was combined with Freebase entities adopted from [42, 125].

[1]https://grouplens.org/datasets/movielens

| Dataset | Domain | User feedback | | Data type | | | Variation | Model |
|---|---|---|---|---|---|---|---|---|
| | | Explicit | Implicit | Metadata | Review | Image | | |
| Movielens | Movie | ✓ | ✓ | ✓ | | | Movielens-HetRec-2011 | [64] |
| | | | | | | | Movielens-100K | [6, 69, 113] |
| | | | | | | | Movielens-1M | [88, 95, 113, 114] |
| | | | | | | | Movielens-20M | [9, 43, 114] |
| Amazon | E-commerce | ✓ | | ✓ | ✓ | ✓ | All categories | [32] |
| | | | | | | | Automotive | [17, 31, 127] |
| | | | | | | | Beauty | [5, 27, 88, 101, 108, 109, 124, 126] |
| | | | | | | | Book | [61, 88, 98, 99] |
| | | | | | | | CDs and Vinyl | [5, 27, 108, 109] |
| | | | | | | | Cell Phones and Accessories | [5, 27, 31, 62, 88, 101, 108, 109, 124, 126, 127] |
| | | | | | | | Clothing Shoes and Jewelry | [5, 27, 88, 101, 108, 109, 124, 126] |
| | | | | | | | Electronics | [62, 69] |
| | | | | | | | Grocery | [31, 127] |
| | | | | | | | Musical Instruments | [17] |
| | | | | | | | Toys and Games | [17] |
| Last.fm | Music | | ✓ | ✓ | | | | [56, 61, 98, 99, 114] |
| Yelp | Business | ✓ | | ✓ | ✓ | ✓ | | [50, 56, 61, 69, 98, 103] |
| Book-Crossing | Book | ✓ | | ✓ | | | | [90, 95, 114] |
| KKBOX | Music | | ✓ | ✓ | | | | [90, 102] |

Table 2. Benchmark datasets used in state-of-the-art explainable graph-based recommender systems

**Amazon**[2] contains product reviews and metadata from Amazon, which include 142.8 million reviews spanning May 1996 to July 2014 from around 20 million users. This dataset includes reviews (containing ratings, textual reviews, and helpfulness votes) and item metadata (descriptions, category information, price, brand, items that were bought/viewed together, and image features). In [32], the whole dataset with items from multiple categories was used to examine the effect of cross-category information on generating explanations. However, the whole dataset can be divided into smaller datasets of different product categories. Each dataset can be used as an individual benchmark, which means that the results obtained from these divided datasets are not necessarily comparable to each other [109]. Some examples of these datasets of various categories, along with the papers in which they were adopted, can be found in Table 2. Similar to the MovieLens datasets, side information from publicly available knowledge bases can be incorporated to construct graphs for recommender systems. In [98] and [61], the graph was constructed by mapping the book titles in the Amazon-Book dataset to the corresponding entities in Freebase [10].

**Last.fm**[3] is a music recommendation dataset extracted from Last.fm online music systems. It contains binary implicit feedback (tagging data) between users and music, as well as social networking and music artist listening information. Also, similar to the Movielens and Amazon datasets, side information from Freebase can be incorporated to construct graphs as in [61].

**Yelp**[4] offers comprehensive information about businesses and users on the Yelp website. It provides user data consisting of information about each user's account, such as the account creation date, the number of reviews posted, the average rating given by the user, and the social network and friendship data among users. As for businesses, this dataset provides information about the business category, location, hours of operation, average rating, number of reviews, and the count of user check-ins by date. In total, there are 1,017 categories of businesses (e.g., "Restaurants", "Bars", etc.) and 6,990,280 reviews over 150,346 businesses.

[2]https://jmcauley.ucsd.edu/data/amazon/
[3]https://grouplens.org/datasets/hetrec-2011/
[4]https://www.yelp.com/dataset/

**Book-Crossing**[5] consists of approximately 1,149,780 book ratings (from 0 to 10) from users in the Book-Crossing community. In [95] and [114], these ratings were converted to implicit feedback. Each user-book interaction was marked as 1 if the user had rated the book, regardless of the rating score, to preserve as much interaction information as possible.

**KKBOX**[6] is a music recommendation dataset provided by a music streaming platform KKBOX. The KKBOX dataset consists of user-song interactions recorded within a specific time duration. For users, the metadata includes user ID, city, age, gender, registration method, registration date, and expiration date. For items (songs), the metadata includes song ID, song length, genre, artist, composer, lyricist, and language. In [90], the entities in the KKBox dataset were mapped to the CN-DBpedia[7] knowledge base [110] to construct the graph.

## 7 EXPLAINABILITY EVALUATION

Evaluating explainability is crucial for the development of explainable graph-based recommender systems. It not only enables comparison between different systems but also guides their refinement and optimization in terms of explainability. This section presents an overview of evaluation techniques used in explainable graph-based recommender systems, followed by a comparative analysis of their strengths and limitations to support the selection of appropriate evaluation methods. However, since explainability in recommendations is a relatively new area, there are not many common metrics for evaluating explanations or explainable recommendations. Some techniques for qualitative and quantitative evaluation are discussed below.

### 7.1 Qualitative Evaluation

Most existing systems evaluate the effectiveness of their explainability via qualitative evaluation. It can be a **visualization** of explanations or an analysis of **case studies** to decide whether their explanations are practical. Moreover, a **user survey** can also be conducted to examine the quality of explanations. In this paper, these three types of qualitative evaluation are discussed. Table 3 summarizes the qualitative evaluation techniques used in some state-of-the-art systems.

**7.1.1 Visualization** Visualization of explanations usually involves the use of graphical elements such as charts, graphs, maps, or other visual aids to illustrate the characteristics of explanations and analyze their effectiveness. The following describes the techniques previously used:

**Heat map**: In [50, 103], interpretability was demonstrated using heat maps of attention weights to highlight important features and identify missing informative features.

**Scatter plot**: In [56], item embeddings were visualized in scatter plots, with each point colored according to its closest associated factor. This enabled the authors to observe item clusters and examine how the model identified factors and learned embeddings, enhancing explainability.

**Explanation paths with attention scores**: In [61], given a randomly selected user-item pair, the explanation paths with high attention scores were extracted to form reasoning sub-graphs. These sub-graphs were displayed to illustrate the explainability performance.

[5]https://www.kaggle.com/datasets/somnambwl/bookcrossing-dataset
[6]https://www.kaggle.com/c/kkbox-music-recommendation-challenge/data
[7]http://kw.fudan.edu.cn/cndbpedia

| Qualitative evaluation method | Advantage | Disadvantage |
|---|---|---|
| Visualization [50], [103], [56], [61] | - Provides a big picture of how the models work and the practicality of the generated explanations<br>- Easy to implement as it does not necessitate additional information or human annotations | - Visualization-based evaluation is subjective, varying based on individual observers' interpretations<br>- More suited for preliminary assessments; additional evaluation techniques are required for better understanding<br>- Some techniques may require user or item sampling, potentially limiting the representativeness of the evaluation |
| Case study [6], [56], [103], [99], [43], [31], [95], [109], [114], [98], [81], [17], [108], [90], [126], [101], [5], [102], [62], [27] | - Allows for in-depth examination of recommendation model explainability by assessing real-world cases of recommendation model explainability<br>- Facilitates a user-centric evaluation, considering the actual impact and usability of the explanations on individuals | - Can be subjective, requiring human interpretation (vary significantly based on prior knowledge and experience)<br>- The uniqueness of each case might not represent the broader user population, making it challenging to draw universal conclusions |
| User survey [9], [6], [88], [127], [61], [32] | - Provides direct insights into users' perceptions of the system's explainability, crucial for improving user satisfaction and trust<br>- Statistical analysis of user responses allows for a more structured and measurable evaluation compared to the other methods | - Typically requires significant expertise and considerable resources<br>- The design of tests or questionnaires is crucial for the outcome. It needs careful consideration, adding complexity to the survey process. |

Table 3. Qualitative methods adopted in state-of-the-art explainable graph-based recommender systems

Visualizing model structures (e.g., attention weights) or generated explanations helps users and developers understand the decision-making logic and assess the usefulness of explanations. It is simple to implement and does not require annotated data. However, visualization is subjective and sometimes relies on sampling, making it suitable mainly for initial evaluations. Further assessment with additional techniques is needed to fully evaluate explainability performance.

**7.1.2 Case Study** A case study typically involves choosing at least one user and showing his/her recommendation obtained from the model, along with its explanation. By comparing the explanation with the user's personal profile or other evidence of their interests, it is possible to evaluate whether the explanation is meaningful and suitable for real-world use. Case studies have been conducted to evaluate various types of explanations, including node and path levels, in various aspects, including practicality, fairness, and diversity. The following discusses the use of case studies in these various scenarios.

**Case studies on practicality of node-level explanations**: In [6], a sampled user and his/her recommendations with explanations were presented to investigate how the model captures this user's semantic attribute preferences. In [56], some entities in a graph with their important scores towards the selected users were presented. In [103], apart from using heat maps, the authors also showed the most important neighbors and the most important multi-style meta-paths of the selected users. In [99], user intents were learned by attentively aggregating information from multiple relations, and case studies showed the relations and their weights forming each intent to reveal which relations influenced a user's preferences. Further, in [100], a case study examined the disentanglement of user intents and their alignment with item semantics. For a selected user, interaction scores were computed for four intents. The highest-scoring item for each intent was analyzed to observe preference variation across different intents and its correspondence with high-level item semantics. In CGSR [117], explainability was assessed by computing similarity scores among items in a user's neighborhood to examine how the model blocked shortcut paths and preserved causal relations. In [106], two session-based recommendation sessions were analyzed, using item-level explanation scores to determine which items most significantly influenced the recommended item and assess their plausibility.

**Case studies on practicality of path-level explanations**: Several path-level explanation models evaluate their effectiveness by extracting explanation paths for sampled users and analyzing their practicality and interpretability. For example, in [43], the authors presented a user's top-3 recommendations along with their explanation paths and discussed their suitability. In [81, 98, 114], paths connecting user-item pairs were extracted using attention weights and examined to reveal user preferences from different perspectives. In [17], item-item paths were retrieved for a user based on learned attention weights and analyzed based on their suitability and practicality. In [90, 101, 108, 126], explanation paths were generated via RL navigation and presented to demonstrate the effectiveness of their explanation-generation methods. In [109], two case studies were presented for selected users, showing each user's layout tree derived from their profile and a subset of explanation paths generated based on that layout tree. In [62], the practicality of explanation rules was verified by first examining whether item pairs connected with these rules correspond with common sense or not. Then, these rules were labeled by three experts to check whether users would agree with these rules. In [106], case studies on two session-based recommendations computed scores reflecting item significance within the session, and the reasonableness of these scores was evaluated.

Besides displaying explanation paths and discussing their practicality, some works included additional assessments. For example, in [95], different explanation paths for a randomly sampled user were compared with each other and the relevance probabilities were analyzed. In [31], case studies were conducted to evaluate performance in addressing recall bias.

Apart from considering explanation paths individually for evaluation, there are some papers that combined such paths to form a sub-graph and conducted case studies on this sub-graph. For example, in [5] and [102], the authors extracted all the explanation paths connecting the selected user-item pair, formed a sub-graph, and examined its high-order connections.

**Case studies on fairness and diversity of path-level explanations**: Apart from the practicality of explanations, case studies can also be conducted to evaluate fairness and diversity. For instance, in [27], the generated explanation paths were compared with those from baseline methods based on the variety of nodes and relations within the paths. Greater diversity in nodes and relations indicates higher fairness, meaning the explanation paths are less biased toward specific nodes or relations. This exemplifies the uses of case studies to evaluate the explainability of graph-based recommender systems in different aspects, not only accuracy or practicality.

Overall, case studies provide a more in-depth, user-centric evaluation of recommendation explainability than visualizations. However, they are subjective, relying on human interpretation, and may not fully evaluate explanation performance. They also have limited generalizability, as individual cases might not represent the broader user population.

#### 7.1.3 User Survey

Conducting a user survey is another straightforward way to evaluate explainability since it can provide users' opinions towards generated explanations or explainable recommendations. This evaluation approach involves recruiting a group of people (which could be stakeholders or potential users of the systems) and using questionnaires/tests to gain knowledge from these users regarding the explainability of recommender systems.

**A/B test**: In [9], the persuasiveness of explanations generated by SemAuto was evaluated through online A/B testing. In the first step, the user had to select at least 15 movies that they have watched from the list provided and rate each movie on a five-star rating scale. These movies were treated as input data for their model. After that, the user was given a list of recommendations based on their selected movies without any explanations. The user was then asked to rate the recommended items. Next, the explanations were presented to the user, and the user was asked to re-rate the

top-2 recommended items. The results before and after providing the explanations were then compared to examine the impact of the explanation in terms of persuasiveness.

**Questionnaire:** In [6], a user survey investigated whether the number of semantic attributes in explanations affected user satisfaction. Thirty-four participants were assigned to low (up to 1 attribute), medium (up to 3), or high (up to 5) groups. Demographic information was collected, and participants rated at least 10 previously watched movies before receiving recommendations with explanations based on their assigned group. Participants then completed a Likert-scale questionnaire on explanation usefulness. Results were analyzed using Analysis of Variance (ANOVA) and Tukey's Honestly Significant Difference (HSD) [1]. In [88], two human evaluations examined (1) whether the generated explanation paths were relevant to highlighted entities, and (2) whether the highlighted entities were sufficient to anticipate a positive item. Participants were asked to rate the relevance and select the next nodes based on highlighted entities until reaching the final positive item. In [127], the authors sampled 50 paths connecting a user to one of his/her previously interacted items in the training set to represent users' historical behaviors. The participants were asked to rank three models based on the consistency between their generated explanation paths and the users' historical behaviors. In [61], 100 user-item pairs were randomly selected, and explanation paths were generated. Ten participants rated the relevance (the likelihood that the explanations are related to the recommended items) and diversity (variety of nodes and relations) of these paths. To reduce variability, the 100 user-item pairs were divided into five groups, with two participants assigned to each group. Final scores were averaged across participants. In [32], counterfactual explanations were evaluated using 500 participants on Amazon Mechanical Turk. For three recommended items, participants saw both a counterfactual explanation and a path-based explanation and answered three questions: "Which method do you find more useful?", "How do you feel about being exposed through explanations to others?" and "Personally, which type of explanation matters to you more".

In summary, user surveys offer direct insight into how users perceive a system's explainability, helping improve satisfaction and acceptance. Statistical analysis of responses allows for structured, measurable evaluation. However, surveys require expertise, resources, and careful design, including defining objectives, selecting target groups, and addressing length, accessibility, anonymity, and privacy. These considerations make user surveys time-consuming and complex to conduct.

### 7.2 Quantitative Evaluation

Quantitative methods use metrics to measure and compare the explainability or suitability of explanations or explainable recommendations. Unlike qualitative methods, they provide concrete, objective results that are not influenced by human judgment. This section reviews quantitative evaluation metrics used in state-of-the-art systems. There metrics are summarized in Table 4.

#### 7.2.1 Mean Explainability Precision (MEP), Mean Explainability Recall (MER) and Mean Explainability F-Score (xF-SCORE)

In [64, 113], the applicability of the models in terms of predicting explainable items was evaluated by two metrics, i.e., Mean Explainability Precision (MEP) and Mean Explainability Recall (MER) [2, 3]. MEP and MER were defined similarly to regular Precision and Recall to measure the accuracy of predicting those items that are explainable to the users. Compared to regular Precision and Recall, instead of considering a set of recommended items and a set of positive items of each user, MEP and MER consider a set of recommended items and a set of explainable items of each user as follows: MEP $= \frac{1}{|\mathcal{U}|} \sum_{u \in \mathcal{U}} \frac{|\mathcal{E}_u \bigcap \mathcal{Y}_u|}{|\mathcal{Y}_u|}$, and MER $= \frac{1}{|\mathcal{U}|} \sum_{u \in \mathcal{U}} \frac{|\mathcal{E}_u \bigcap \mathcal{Y}_u|}{|\mathcal{E}_u|}$, where $\mathcal{U}$ denotes the users set, $\mathcal{E}_u$ denotes the set of explainable items of user $u$ and $\mathcal{Y}_u$ denotes the set of recommended items of user $u$.

| Quantitative evaluation method | Advantage | Disadvantage |
|---|---|---|
| Mean Explainability Precision (MEP), Mean Explainability Recall (MER), Mean Explainability F-Score (xF-SCORE) [113], [6], [64] [113], [6], [64] [6] | - Evaluate implicit explanations<br>- Provide a nuanced understanding of model performance, such as detailed misclassification comprehension<br>- No additional user/item information is required | - Rely on having a set of explainable items for each user, introducing complexities when identifying explainable items for each user<br>- Not originally designed for evaluating explicit explanations generated by the model |
| Performance shift [113], [56] | - No additional user/item information is required<br>- Evaluates explicit explanations<br>- Provides valuable insights into the operational dynamics of the model, contributing to the transparency of recommender systems | - Rely on the chosen evaluation metric, such as Recall shift, which may not capture all relevant aspects of model performance<br>- May oversimplify the nuanced impact of certain nodes and relations on the recommendation model, potentially leading to misinterpretations |
| Simpson's Index of Diversity (SID) [27], [88] | - Offers another aspect of explanation quality that should be evaluated besides accuracy<br>- No additional user/item information is required | - Other evaluation metrics focusing on accuracy should be jointly considered to ensure both accuracy and diversity of explanations<br>- Only path-level explanations are applicable |
| Review matching [124], [88] | - Enables the assessment of the alignment between the generated explanations and the characteristics of users and items<br>- Ensures explanation quality, potentially enhancing user satisfaction with recommender systems | - Requires users' reviews, limiting its applicability in cases where such information is unavailable<br>- Previously used for the assessment of path-level explanations, modifications are needed to extend its application to other types of explanations |
| Jensen–Shannon (JS) divergence [127] | - Enables the assessment of how well the generated explanations align with users' historical behaviors (similar to review matching)<br>- Requires only user-item paths extracted from a graph (does not need users' reviews) | - Extracting user-item paths is computationally expensive, limiting consideration to subsets of paths<br>- Determining the optimal number of sampled paths is challenging: too few may inadequately capture a user's history, while too many can introduce noise |
| Levenshtein distance [61] | - Evaluates the quality of the generated explanations by comparing them with ground-truth explanations or compares the similarity of explanations generated via different methods<br>- No additional user/item information is required<br>- Computationally efficient using dynamic programming | - If there are no ground-truth explanations, the generated explanations can only be compared with other generated explanations<br>- Only path-level explanations are applicable<br>- Sensitive to sequence length<br>- Originally designed for sequences, e.g., strings; for subgraphs, metrics such as graph edit distance [18] may be more suitable |

Table 4. Quantitative evaluation methods in state-of-the-art explainable graph-based recommender systems

In [6], in addition to MEP and MER, Mean Explainability F-Score (xF-SCORE) was also used. This xF-SCORE can be computed by MEP and MER as xF-SCORE = $2 \cdot \frac{\text{MEP}\cdot\text{MER}}{\text{MEP}+\text{MER}}$

MEP, MER, and xF-SCORE offer a nuanced assessment of model performance beyond accuracy alone, without needing additional user or item data such as reviews or descriptions. However, calculating these metrics relies on having a set of explainable items for each user in the dataset. To obtain a set of explainable items of each user, different strategies can be used. For instance, it can be a set of items with at least one explanation generated by a post-hoc explaining method [113] or a set of items whose explainability values/scores are higher than a pre-defined threshold [64]. This process adds complexity to evaluation and demands setting specific thresholds, which can be challenging without prior expertise. Finally, these metrics were employed to assess how effectively a model generates explainable items, rather than explicitly evaluating the quality of explanations themselves. Adapting these metrics for the evaluation of generated explanations may require further refinement.

**7.2.2 Performance Shift** Performance shift involves evaluating changes in predictive outcomes resulting from the modification or removal of specific elements within a model or its input. It assumes that elements are good explanations if altering them notably affects predictions. In [113], the shift in their model's performance was examined to determine the significance of nodes and relations within the generated explanation paths. Specifically, for a given user-item pair and its explanation path, the nodes and relations within this path were excluded from the training data. If re-training the model with this modified dataset caused a significant drop in the predicted rating, it indicated that the removed elements were important, suggesting the explanation path captured key information about the model's operation. In [56], the generated explanation in the form of significant concepts (or attributes) was evaluated by adversarial perturbation [26] similarly to [113]. The evaluation involved sampling 200 users from the test set, predicting recommendations, and identifying the important item or attribute nodes for each user-recommended item pair. By removing these critical nodes, new recommendations were generated and compared with the original ones. The Recall shift, indicating the difference in Recall values between the new and original recommendations, was then calculated. A higher shift value suggests that the explanations accurately and substantially reflect users' preferences.

Evaluating explainability using performance shift does not require auxiliary user and item information, similar to MEP, MER, and xF-SCORE. However, unlike these metrics, performance shift is used to evaluate explicit explanations. In addition to assessing explanations, analyzing performance shift provides insights into the operational dynamics of the model, contributing to the transparency of recommender systems. Nevertheless, the effectiveness of performance shift analysis heavily depends on the chosen evaluation metric (e.g., Recall shift), which may not capture all aspects of model performance. Moreover, within a recommendation model, nodes and relations are interconnected, and their effects are intertwined. Assuming that a change in rating score indicates significance may oversimplify the nuanced impact of specific nodes and relations on the model.

**7.2.3 Simpson's Index of Diversity (SID)** Current explainable recommender systems may face bias issues, potentially discriminating between users when providing explanations [27]. Thus, fairness has become an important evaluation criterion. One way to assess fairness is by measuring the diversity of generated explanations. Low diversity may indicate that the model is biased toward certain explanations, reducing fairness. In [27, 88], Simpson's Index of Diversity (SID) [82] was adopted to quantify the diversity of generated explanation paths. Specifically, it was used to measure the probability that two randomly selected explanation paths belong to the same path pattern (or meta-path) computed as follows: $\text{SID}(\Pi, M) = 1 - \frac{\sum_{i=1}^{|\Pi|} m_i(m_i-1)}{M(M-1)}$ where $\Pi$ is the set of all path patterns (or meta-paths), $m_i$ denotes the number of paths belonging to the $i$th pattern, and $M$ is the total number of user-item paths. Ranging from 0 to 1, the higher the SID is, the more diverse the explanation paths. Based on SID, FairKG4Rec [27] was compared with the other baselines in terms of diversity and fairness in recommendations and explanation paths. SID was also adopted in [88] to quantify the diversity of explanation paths obtained from UCPR.

SID offers another aspect of explanation quality besides accuracy and can be computed without any additional user/item information. However, while a model may produce diverse explanation paths for a user, they might not accurately match the user's characteristics or interests. Hence, other accuracy-focused evaluation metrics should also be considered to ensure both accuracy and diversity of explanations. Additionally, as SID assesses path-level explanations only, adaptations are necessary to evaluate the diversity of other explanation types.

**7.2.4 Review Matching** Review matching is based on the assumption that the quality of explanations is positively correlated with the degree of alignment between the explanations and users' reviews. For instance, in [124], the ground-truth reviews were used to evaluate the generated explanation paths. For each positive recommended item, words in its review were filtered based on their frequencies and TF-IDF values, leaving the remaining words as ground-truth reviews. Subsequently, the nodes in the generated explanation path for that item were gathered and ranked according to their frequencies. The explainability of the path was evaluated by comparing its nodes with the ground-truth words, using Precision@K and Recall@K metrics based on the top-5 matched nodes. These metrics can be computed as follows:

Precision@K = $\frac{|\mathcal{V}_{p_i} \cap \mathcal{V}_{r_i}|}{|\mathcal{V}_{p_i}|}$, and Recall@K = $\frac{|\mathcal{V}_{p_i} \cap \mathcal{V}_{r_i}|}{|\mathcal{V}_{r_i}|}$, where $p_i$ denotes an explanation path, $r_i$ denotes the item review associated with path $p_i$, $\mathcal{V}_{p_i}$ is the set of selected top-K nodes in path $p_i$, and $\mathcal{V}_{r_i}$ is the set of ground-truth words in review $r_i$. Similarly, UCPR [88] also evaluated the explanation paths discovered by the RL agent by conducting a review matching. Recall@K and NDCG@K were applied based on the top-10 matched entities. NDCG@K can be computed as follows: NDCG@K = $\frac{\text{DCG@K}}{\text{IDCG@K}}$, where DCG@K = $\sum_{i=1}^{K} \frac{\text{rel}_i}{\log_2(i+1)}$ and IDCG@K = $\sum_{i=1}^{K} \frac{1}{\log_2(i+1)}$ where $\text{rel}_i$ is the relevance of the node in the explanation path at rank $i$.

Review matching assesses how well generated explanations align with users' interests, improving explanation quality and potentially increasing user satisfaction. However, it requires access to user reviews, limiting its use in datasets without them. Moreover, previous studies mostly applied review matching to assess path-level explanations, requiring adjustments for other explanation types.

**7.2.5 Jensen–Shannon (JS) Divergence** Faithfulness is one of the principles in explaining machine learning and deep learning models [15, 60]. It focuses on how well the generated explanations accurately reflect the behavior or decision-making process of the underlying model [15]. In the context of recommendation systems, a faithful explanation should be personalized and capable of accurately reflecting a user's historical behavior. In [127], inspired by [80, 83, 93], the Jensen–Shannon (JS) divergence of rule-related distributions in the training set and the set of generated explanations was adopted to measure this quality. For each user $u$, let $F(u)$ denote the distribution of rules (defined similarly to meta-paths) over the user-item paths sampled from a training set, $Q_f(u)$ denote the distribution of rules over generated explanation paths, and $Q_w(u)$ denote the distribution of rule weights derived from either rule importance scores or path weights of the explanation paths. The JS scores were computed by $\text{JS}_f = \mathbb{E}_{u \sim U}[D_{\text{JS}}(Q_f(u)||F(u))]$ and $\text{JS}_w = \mathbb{E}_{u \sim U}[D_{\text{JS}}(Q_w(u)||F(u))]$, where $D_{\text{JS}}$ denotes the JS divergence. Smaller values of these scores indicated better faithfulness of the generated explanation paths.

Similar to review matching, JS divergence evaluates how the generated explanations align with users' historical behaviors. However, JS divergence relies on user-item paths extracted from a graph, enabling evaluation even without users' reviews. Nevertheless, extracting user-item paths can be computationally expensive [51]. Due to this time-consuming nature, computing JS divergence may necessitate considering only a subset of users and user-item paths, as in [127]. Moreover, the number of sampled paths can significantly affect the evaluation. A small number may inadequately represent a user's historical behavior, while a large number could introduce noise [85, 95].

**7.2.6 Levenshtein Distance** In [61], explanation sub-graphs generated from their proposed model KR-GCN and the other baselines were compared in terms of their topological similarity by using Levenshtein distance [67]. Levenshtein distance is a metric used to quantify the difference between two sequences. It measures the minimum number of edits (insertions, deletions, or substitutions) required to transform one sequence into another. It is computed using dynamic programming, defined by the following recurrence relation:

$$\text{lev}(a, b) = \begin{cases} 0 & \text{if } \min(|a|, |b|) = 0 \\ \text{lev}(a[2:], b[2:]) & \text{if } a[1] = b[1] \\ 1 + \min(\text{lev}(a[2:], b), \text{lev}(a, b[2:]), \text{lev}(a[2:], b[2:])) & \text{otherwise} \end{cases} \quad (1)$$

where $\text{lev}(a, b)$ is the Levenshtein distance between sequences $a$ and $b$, $a[1]$ denotes the first element of sequence $a$, $a[2:]$ denotes the subsequence of $a$ excluding its first element, and $|a|$ denotes the length of sequence $a$.

Levenshtein distance enables comparison of explanations, including paths or sub-graphs, without requiring additional user/item information. It can be efficiently implemented using dynamic programming. The quality of generated explanations can be assessed by comparing them with ground-truth explanations, though these may be limited. As in [61], the assessment of generated explanations only involved comparing them with other generated explanations from baseline models, focusing on their similarity rather than the alignment with ground-truth explanations. Levenshtein distance can be sensitive to sequence length, with longer sequences resulting in higher distances. Moreover, it was originally designed for comparing sequences such as strings. For sub-graphs, other metrics, e.g., graph edit distance [18], might be more appropriate.

## 8 CONCLUSIONS AND FUTURE DIRECTIONS

### 8.1 Conclusions

In this paper, state-of-the-art explainable graph-based recommender systems were summarized and discussed. Such systems were categorized based on three aspects, i.e., learning methods, explaining methods, and explanation types. Based on learning methods, explainable graph-based recommender systems can be divided into three approaches: an *embedding-based approach* using node embedding methods for learning recommendations, a *path-based approach* that extracts actual paths containing high-order information and feeds them to a recommendation framework, and a *hybrid approach* that combines both embedding-based and path-based approaches. Compared to path-based methods, embedding-based methods are more flexible because node embeddings capture the overall graph structure and can be adapted to different recommendation frameworks. Path-based methods, on the other hand, explicitly leverage multi-hop relations, making path-level explanations easier to generate. However, pre-extracting paths can be time-consuming, and poorly chosen paths may introduce noise that harms recommendation performance.

Based on explaining methods, explainable graph-based recommender systems can be divided into two approaches: a *model-specific approach* and a *model-agnostic approach*. Model-specific methods often use attention mechanisms or RL-based path reasoning to generate explanations. Auto-regressive path generation is less common because its generated paths may not exist in the graph, making it less reliable. Model-agnostic methods instead rely on criteria or metrics to rank nodes, relations, or paths and select the most suitable ones as explanations. This provides flexibility but can be computationally expensive, as candidate extraction and validation are time-consuming. Pre-processing or candidate filtering can mitigate this. However, post-hoc explanations can be misleading since they are generated separately from the model and may not truly represent the model's reasoning. Low-quality data or features can also further reduce their usefulness.

Lastly, based on the types of explanations, explainable graph-based recommender systems can be divided into four approaches: *node-level* type (e.g., important nodes), *path-level* (e.g., explanation paths), *meta-path-level* (e.g., significant meta-paths), and *implicit* type (the models are constrained to produce more explainable recommendations than those non-explainable ones). Path-level explanations are more ubiquitous than the other types. Compared to node-level explanations, path explanations consist of more details, which can be more useful when presenting to users. Also, compared to meta-path level explanations, path-level explanations are more specific since entities (nodes and relations) are presented instead of node types and relation types.

### 8.2 Future Directions

Despite significant progress in developing explainable graph-based recommender systems in recent years, several gaps and challenges remain. This section summarizes key challenges and potential directions for future research.

**Explainability via attention mechanisms** Although attention mechanisms can highlight important graph components as explanations, they can be misleading. A study in Natural Language Processing (NLP) [45] showed that attention weights do not always align with gradient-based measures of feature importance. A model may produce the same output even when attending to different features, so it is not always accurate to conclude that specific attended features drive a particular prediction. Nevertheless, there is an argument in [105] saying that attention weights could be used as explanations depending on the definition of explainability. The authors suggested that attention weights may serve as *plausible* explanations but may not be faithful enough to indicate the model's decision-making logic or the exact correlation between inputs and outputs. Despite differences between NLP and recommendation tasks, these studies highlight the need to carefully examine how attention weights truly contribute to explaining recommendations.

**Limitation of graph-based explanations** As pointed out in [56], if the stakeholders are end users and the goal of explainability is to persuade them, importance scores or attention weights may not be suitable in that case. To be specific, identifying significant nodes/relations/paths/meta-paths may be more useful for examining the model structure or debugging the model rather than persuading users. Also, it has been shown in Section 4 that path-level explanations are quite popular in explainable graph-based recommender systems. One reason is that it is intuitive to leverage paths carrying high-order information from graphs for generating explanations. However, a path extracted from a graph may not serve as a user-friendly explanation, especially when the goal is to increase users' comprehension of the recommendation. A previous effort has been made to provide explanations in a form that is more human-language-like to ensure users' understanding [5]. Moreover, prior efforts mostly focused on extracting paths that connect the user with the recommended item. Such paths typically contain other users' actions, which raise some privacy issues [32]. Therefore, future directions in this area could be further exploration of different types of explanations that are easily comprehensible for users and also preserve users' privacy.

**Multimodal graphs** A multimodal graph typically combines data from different sources, such as text, images, and audio, to create a more comprehensive user and item representation for learning recommendations. By using multiple types of information, a multimodal graph can provide more accurate and diverse recommendations to users [43, 44, 63, 64, 84, 104]. Additionally, since a multimodal graph considers different types of data, it can produce recommendations based on a wider range of factors, such as user preferences, previous behavior, and other context-specific information. This can lead to more personalized and relevant recommendations, which can improve the overall user experience. Also, it can improve explainability since broader factors are considered. However, most previous work typically ignored multimodal data and focused only on user and item metadata. Thus, effectively incorporating multimodal data in explainable graph-based recommender systems is still an open research question.

**Causal graphs** In the research area of explainable recommender systems, causal reasoning is one of the popular approaches along with graph-based approaches [29]. These systems, based on causal reasoning, aim to enhance transparency and interpretability in the recommendation process by incorporating causal relationships between various factors [29]. Specifically, they utilize causal graphs to visualize and comprehend cause-effect relationships between factors such as users, items, and attributes, drawing assumptions for learning recommendations [14, 111, 112]. Unlike other graph-based recommender systems, they do not emphasize leveraging connectivity information or high-order relations between different entities within graphs. However, some state-of-the-art models have successfully integrated causal relations into graphs to simultaneously harness high-order connectivity information and causal reasoning [100, 106, 117]. This showcases the potential of combining graphs, particularly knowledge graphs or information networks, with causal graphs to incorporate causal reasoning alongside high-order connectivity information. Nevertheless, there is still room for improvement in the seamless integration of causal relationships into graphs, aiming for more effective and nuanced models. Constructing causal graphs that specifically focus on causal multi-hop relations represents a promising avenue for advancement. Overall, the development of graph-based explainable recommender systems that are capable of considering multi-hop relations and causal relations simultaneously will open up new opportunities for explainable recommendations.

**Scalability** Graphs generated from real-world data typically contain a large number of nodes and relations. As a result, it is challenging to apply recommender systems on these graphs due to the number of computational resources required [52, 53]. Moreover, despite the fact that augmenting graphs with auxiliary information can increase the accuracy and explainability [54, 63, 101], it also increases the number of nodes and relations and computational complexity. All of these challenges cause scalability issues for numerous graph-based recommender systems. Many approaches have been proposed to deal with such issues. One of these approaches is to reduce the size of a graph, for example, compressing relations in a graph to reduce its size [81] and pruning relations to reduce the size of action spaces in RL-based path reasoning models [90, 108]. Another approach is using sub-graphs [25] or subsets of paths [36, 115] instead of using a whole graph or all paths which is not scalable. Apart from the input side, scalable methods for extracting and leveraging high-order information were also proposed to cope with the scalability issues [38, 64, 123]. Although these approaches can attenuate the scalability issues, the accuracy and explainability performances could be compromised since certain parts in a graph are disregarded. Therefore, how to develop a scalable graph-based recommender system while maintaining high accuracy and explainability performances still needs more investigation.

**Benchmark datasets and evaluation** Benchmark datasets can facilitate testing and comparing the performance of different models. As previously mentioned, benchmark datasets for evaluating the explainability of graph-based recommender systems are not commonly available. Due to the lack of benchmark datasets, validating the effectiveness of each model in terms of explainability becomes difficult. Also, compared to other performance aspects such as accuracy, novelty, and diversity of recommendations, there has been no conventional or universal method/metric for explainability. Some previous studies employed highly specific methods to evaluate their models' explainability. Consequently, comparing these models with others is challenging, as the evaluation methods may not be applicable across different models. Thus, having benchmark datasets and conventional evaluation metrics could be one of the key factors to further development of effective explainable graph-based recommender systems.

**Practical limitations in real-world deployments** Explainable graph-based recommender systems, as well as other recommender systems, face several challenges in real-world deployments. First, continuous training is necessary to accommodate evolving user preferences [96]. This adaptation often requires dynamic graph learning techniques [89, 97, 119]. However, explainable recommender systems based on dynamic graphs remain underexplored [97]. A more recent concept for handling continuously evolving graphs and addressing the catastrophic forgetting problem in dynamic graph learning is continual graph learning (CGL) [120, 121]. Incorporating CGL into explainable graph-based recommender systems represents a promising research direction to enhance performance, enabling systems to adapt seamlessly to changing user–item interactions. Ensuring data privacy is another critical challenge. Explainable graph-based recommender systems often leverage user connectivity information, which can reveal sensitive interactions. This results in the need for privacy-preserving techniques and strict data handling in such systems [30]. Finally, offline metrics typically fail to simulate actual user contexts, interaction feedback, and evolving data distributions. This can lead to deployment scenarios where recommendation quality and explainability diverge sharply from test results [47]. Developing explainable graph-based recommender systems that perform well both offline and in real-world online environments remains a challenge.